%% file: arxiv_main.tex
\documentclass[11pt]{article}

\usepackage[numbers]{natbib}
\usepackage{arxiv}
\usepackage{wrapfig}
\usepackage{booktabs}       
\usepackage{amsfonts}       
\usepackage{amsmath}
\usepackage{amsthm}
\usepackage{amssymb}
\usepackage[colorlinks,linkcolor=magenta,filecolor=blue,citecolor=blue,urlcolor=blue]{hyperref}%
\usepackage[capitalize,noabbrev]{cleveref} 
\usepackage{graphicx}
\usepackage{subcaption}
\usepackage{algorithmic}
\usepackage[vlined,linesnumbered,ruled,resetcount]{algorithm2e}
\usepackage{wrapfig} 
\usepackage{multirow}
\usepackage{url}

\usepackage{circledsteps}
\usepackage{pifont}
\usepackage{tikz}

\usepackage{xcolor}         
\usepackage{pifont}
\definecolor{redx}{RGB}{180,0,0}
\definecolor{greenx}{RGB}{0,180,0}

\definecolor{redx}{RGB}{180,0,0}
\definecolor{greenx}{RGB}{0,180,0}
\usepackage{threeparttable}

\theoremstyle{plain}

\theoremstyle{definition}

\theoremstyle{remark}

\title{Bridging Today and the Future of Humanity: \\ AI Safety in 2024 and Beyond}

\author{
 Shanshan Han\thanks{
The idea for this article struck the solo author unexpectedly on an ordinary afternoon as she moved into a garage in Palo Alto during the summer of 2024.
} \\
  University of California, Irvine\\
  \texttt{shanshan.han@uci.edu} 
}

\begin{document}

\maketitle

\begin{abstract}
\input{chapters/0_abstract}
\end{abstract}

\input{chapters/1_intro}

\input{chapters/2_related_works}

\input{chapters/3_from_the_new_world}

\input{chapters/4_bridging_today_and_future}

\input{chapters/5_back_to_2020s}

\input{chapters/7_ai_safety_practitioners_in_2020s}

\input{chapters/8_conclusion}

\bibliographystyle{plain}
\bibliography{reference}

\end{document}

%% file: chapters/0_abstract.tex
The advancements in generative AI inevitably raise concerns about their risks and safety implications, which, in return, catalyzes significant progress in AI safety.
However, as this field continues to evolve, a critical question arises: are our current efforts on AI safety aligned with the advancements of AI as well as the long-term goal of human civilization? This paper presents a blueprint for an advanced human society and leverages this vision to guide current AI safety efforts. 
It outlines a future where the \textit{Internet of Everything} becomes reality, and creates a roadmap of significant technological advancements towards this envisioned future.
For each stage of the advancements, this paper forecasts potential AI safety issues that humanity may face. By projecting current efforts against this blueprint, this paper examines the alignment between the current efforts and the long-term needs, and highlights unique challenges and missions that demand increasing attention from AI safety practitioners in the 2020s. 
This vision paper aims to offer a broader perspective on AI safety, emphasizing that our current efforts should not only address immediate concerns but also anticipate potential risks in the expanding AI landscape, thereby promoting a safe and sustainable future of AI and human civilization.

%% file: chapters/1_intro.tex
\section{Introduction}

The rapid developments of AI and Large Language Models (LLMs) have fostered extensive progress in AI safety~\citep{chua2024ai,du2024mogu,hubinger2024sleeper,murule,peng2024navigating,tedeschi2024alert}. Researchers have been dedicated to addressing potential safety risks in AI lifecycle, aiming at aligning AI behaviors with human values and preventing inappropriate model outputs, information leakage, misuses of AI models, etc. However, despite the siginificant efforts on AI safety, a critical question emerges: 
are our current efforts aligned with the advancements of AI and the long-term goal of human civilization, or are they simply addressing the immediate concerns of the 2020s?

One fundamental reason for this uncertainty lies in the probabilistic nature of AI~\citep{Neal2006PatternRA}. Despite their impressive capabilities in natural language processing and problem-solving~\citep{yang2022large,xu2022systematic,huzhang2021aliexpress,duong2024analysis,clusmann2023future}, today's AI, including advanced LLMs~\citep{zhao2023survey,wu2023bloomberggpt,thirunavukarasu2023large}, falls short of what could be considered as ``genuine intelligence''. 
Current AI models rely heavily on vast training datasets to function effectively, yet lack consciousness, self-awareness, and real reasoning abilities comparable to human cognition. They are, in essence, highly sophisticated pattern recognition and prediction machines, rather than entities with authentic logical capabilities.
Recent studies argued that the reasoning abilities of AI models might be a form of approximate retrieval and deductive closure of the training data~\citep{llm_fake_reasoning_ability,Valmeekam2024LLMsSC}. While this process can simulate deductive reasoning in simpler cases through external validation, optimization, and repeated searching of the problem space, 
it differs fundamentally from human reasoning that involves abstract thinking, causal understanding, generalizing from limited examples, etc.

The energy issue is another critical factor that challenges AI as well as our efforts on AI safety~\citep{ahmad2021artificial}. Current AI fails to represent the third industrial revolution from the perspective of human history, as the fundamental energy issue remains unsolved. Historically, the two industrial revolutions that have shaped human civilization were driven by revolutionary energy innovations~\citep{groumpos2021critical}, with the First Industrial Revolution fueled by steam and coal, and the Second Industrial Revolution characterized by technological innovations powered by electricity and petroleum~\citep{groumpos2021critical}.
However, current AI, rather than solving energy issues, consumes a significant amount of energy. Training 
GPT-4 consumed over 50,000 MWh, 10,353.5 tons of CO$_2$ equivalent, and approximately 0.02\% of California's annual electricity generation~\citep{argerich2024measuring,AI_energy_issue,patterson2021carbon}. Inference with LLMs is computationally intense as well, \textit{e}.\textit{g}., a single query to GPT-4 consumes 0.001 to 0.01 kWh, approximately 15x energy than a Google query~\citep{inference_energy,inference_energy2}. Given that ChatGPT has over 200 million weekly active users and receives over 1.54 billion page visits monthly~\citep{gpt_everyday_visit}, when scaled to billions of queries, the energy consumption becomes substantial.
The significant energy consumption of AI raises concerns about their 
long-term sustainability, posing significant challenges to their widespread deployment and scalability. Thus, 
until humanity solves the energy issues, the potential of AI to reshape society will remain limited and uncertain.

\begin{figure*}
    \centering
    \includegraphics[width=0.98\linewidth]{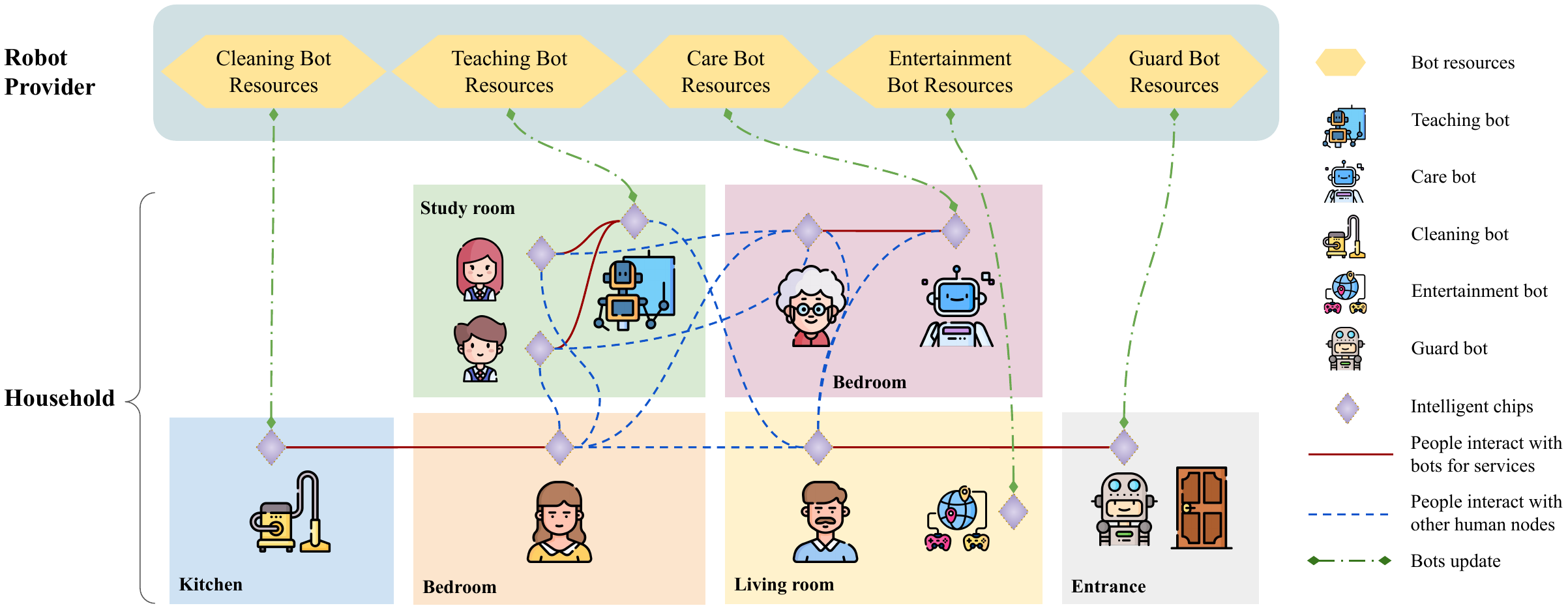}
    \caption{Illustration of Internet of Everything in a Household}
    \label{fig:household-internet-of-everything}
\end{figure*}

These concerns indicate that our current efforts on AI safety might overlook significant issues that might arise as AI continues to evolve. We potentially fail to capture deeper insights that might ultimately shape the impact of AI on human society.
This raises several questions:

\begin{itemize}
    \item Are our efforts on AI safety align with the long-term advancements of AI? 
    \item Are we potentially overlooking significant challenges that will emerge as AI evolves?
    \item Are we addressing real challenges that align with the development of human civilization?
    
\end{itemize}

\underline{\textit{``The Moon and Sixpence.''}} This paper envisions an ultimate advanced society in a distant future of human civilization, and anticipates potential technological advancements to guide today's AI safety efforts. 
In this advanced future, energy issues have been resolved, and breakthroughs in foundational theories, such as Einstein’s relativity theories~\citep{einstein1922general,einstein2013principle} and quantum mechanics~\citep{einstein1935can,zettili2009quantum}, have catalyzed revolutionary changes in AI and human society.
Intelligent chips~\citep{hsu2020intelligent,li2023electromagnetic} and brain-computer interfaces~\citep{he2020brain,nicolas2012brain,wolpaw2013brain} have been fully developed, and advanced intelligent robots are integrated into everyday life. Humans and robots are connected in an interconnected network, \textit{i}.\textit{e}., the \textit{Internet of Everything}; see \cref{fig:household-internet-of-everything}. 
While this vision may be beyond the horizon of several decades, examining the present landscape of AI safety through the lens of this long-term perspective offers valuable insights into the potential and limitations of our current efforts on AI safety. It allows us to identify specific missions for AI safety researchers and practitioners in the 2020s. 
It also reminds us that, while searching for sixpence scattered on the ground and addressing today's immediate concerns, we should avoid being limited by short-term solutions, and instead, lift our gaze and seek the moon. 

\underline{\textit{Takeaways: }} 
\textit{i}) The term ``AI safety'' can be extended to ``\textit{\textbf{AI quality assurance}}'' that encompasses more comprehensive concepts and possibilities in the fast evolving AI landscape. 
\textit{ii}) Deeper collaborations between experts from different fields 
are essential to align AI with human values better.
\textit{iii}) 
AI and AI safety workflow customizations are crucial to ensure that AI systems can adapt to diverse user requirements while maintaining safety and reliability across various use cases.
\textit{iv})
Exploring and understanding interactions between users and AI are crucial for enhancing user experiences, especially in dynamic contexts across different user groups.


%% file: chapters/2_related_works.tex
\input{src/related_work_table}

\input{src/benchmark_table}

\section{Related Works}

In recent years, numerous AI safety surveys have been published~\citep{kenthapadi2024grounding,ayyamperumal2024current,liu2024exploring,chua2024ai,Rahman2023SecurityRA,liu2023trustworthy,he2024security,yao2024survey}, focusing on different aspects of this field, such as jailbreaking, red teaming, ethics, etc~\citep{liu2023trustworthy,yao2024survey,mozes2023use}. Table~\ref{tab:ai_safety_papers} summarizes academia efforts, Table~\ref{tab:ai_safety_industry} summarizes industry insights, and Table~\ref{tab:benchmarks} summarizes AI safety benchmarks.

Among these works, \cite{liu2023trustworthy} 
provides comprehensive insights into building trustworthy AI systems by focusing on reliability, safety, fairness, robustness, and adherence to social norms.
\cite{chua2024ai}  addresses AI safety concerns from the perspective of different entities, \textit{e}.\textit{g}., data and prompts, in LLM lifecycle, and discusses data safety, model safety, prompt safety, alignment, and the complexities of scaling.
~\cite{kenthapadi2024grounding} provides valuable enterprise insights into the practical challenges and opportunities for AI safety in industries.
\cite{gabriel2024ethics} provides a systematic analysis of the ethical risks of advanced AI assistants, and discusses their potential influence on users, risks of misuse, strategies for enhancing human-AI interactions, and the broader implications for human society.
\cite{dalrymple2024towards} proposes the concept of ``Guaranteed Safe'' AI, emphasizing the need for formal, verifiable, and auditable safety guarantees to ensure robust AI behavior. 
\cite{bajcsy2024human} 
proposes a technical roadmap towards next-generation human-centered AI safety that aims to anticipate and prevent potential hazards in the interactions between AI systems and users. 
It emphasizes leveraging control-theoretic safety methodologies alongside the rich interaction models of generative AI, suggesting that a deeper understanding of the feedback loop between AI outputs and human behavior is essential for establishing robust and meaningful safety assurances.

Different from existing works, this paper envisions an advanced society in the future, and use this vision to guide current AI safety efforts. By aligning the current landscape of AI safety with this distant future, this paper discusses the potential and limitations of today's efforts, and identifies special missions of AI safety researchers in the 2020s. This paper encourages maintaining a forward-looking vision while resolving immediate concerns, such that we can ensure that today’s solutions are aligned with the long-term evolution of AI and human society.

%% file: src/related_work_table.tex
\begin{table*}[t]
\centering
\begin{tabular}{@{}p{4cm}|p{1cm}|p{2.5cm}|p{8cm}@{}}
\toprule

\textbf{Paper} & \textbf{Type} & \textbf{Topic} &\textbf{Notes} \\\midrule
``Current state of risks''~\citep{ayyamperumal2024current} & Survey & Guardrails & Guardrails and their implementation challenges \\
\midrule
``Exploring Vulnerabilities and Protections''~\citep{liu2024exploring} & \multirow{2}{16em}{Survey} & \multirow{2}{16em}{Vulnerabilities} & Vulnerabilities of LLMs, especially prompt injections and  backdoor; defenses against the attacks\\
\midrule
\multirow{2}{16em}{``AI safety in LLMs''~\citep{chua2024ai}} & \multirow{2}{5em}{Survey} & Safety issues in 
LLM lifecycle & Safety issues towards entities in LLM lifecycle; Data/model/prompt safety; alignment; safety at scale \\
\midrule
``Security risk''~\citep{Rahman2023SecurityRA} & Survey & Attacks & Attacks and risks in AI models; defensive methods\\
\midrule
``Security of AI Agents''~\citep{he2024security} & \multirow{2}{16em}{Survey} &\multirow{2}{16em}{Agent security} & 
\multirow{2}{25em}{Security issues, vulnerabilities, and defenses for agents }
\\
\midrule
``The good, the bad, the ugly''~\citep{yao2024survey} & \multirow{2}{16em}{Survey} & Risks; vulnerabilities; defenses & LLMs as tools for security and privacy; potential risks and threats; vulnerabilities and defenses. \\\midrule
\multirow{2}{10em}{``Use of LLMs''~\citep{mozes2023use}} & \multirow{2}{16em}{Survey}& Vulnerabilities, threats, defenses,  &  Prevention measures, vulnerabilities, and methods to mitigate the risks associated with the misuse of LLMs
\\

\midrule

\multirow{3}{16em}{``AI risk management''~\citep{qi2024ai}} &\multirow{3}{16em}{Insight} & \multirow{3}{10em}{Safety; security; governance} & 
Bridging AI safety and security in risk management; introducing a reference framework to facilitate common understanding of their differences and interplay.
\\
\midrule
\multirow{3}{16em}{``Human-ai safety''~\citep{bajcsy2024human}} & \multirow{3}{16em}{Insight} & \multirow{3}{9em}{AI and control systems safety} & A safety framework for human–AI interactions; a roadmap towards next-generation human-centered AI safety \\
\midrule
\multirow{3}{16em}{``Managing AI risks''~\citep{bengio2024managing} }& \multirow{3}{16em}{Insight} & \multirow{3}{16em}{Governance} &Emphasizing urgent needs for a balanced approach combining technical research and proactive governance to manage risks of AI systems\\
\midrule
\multirow{2}{16em}{``Guaranteed safe AI''~\citep{dalrymple2024towards}} & \multirow{2}{16em}{Position} &  \multirow{2}{16em}{Framework} & Define Guaranteed Safe AI to equip AI systems with formal, verifiable and auditable safety guarantees. \\

\bottomrule
\end{tabular}
\caption{Overview of AI Safety Papers}
\label{tab:ai_safety_papers}
\end{table*}

\begin{table*}[t]
\centering
\begin{tabular}{@{}p{4cm}|p{1cm}|p{2.5cm}|p{8cm}@{}}
\toprule

\textbf{Paper} & \textbf{Type} & \textbf{Topic} &\textbf{Notes} \\\midrule

``Grounding and Evaluation for LLMs''~\citep{kenthapadi2024grounding} & \multirow{2}{16em}{Survey} & \multirow{2}{16em}{Grounding; safety} & \multirow{2}{20em}{Offer valuable insights for enterprises }\\
\midrule
ByteDance alignment~\citep{liu2023trustworthy} & Survey & Broad concepts & Reliability, fairness, explainability, etc.\\


\midrule
DeepMind ethics paper~\citep{gabriel2024ethics} & Vision & AI ethics & Ethical challenges; 
interactions; impacts on society  \\\midrule
DeepMind evaluation~\citep{weidinger2024holistic} & Insight & Safety evaluation & Safety evaluation for genAI systems at DeepMind
\\\midrule
\multirow{3}{12em}{OpenAI Safety Practices~\citep{openai_lessons_learned,openai_reimaging,openai_preparedness,openai_safety_update}} & \multirow{3}{16em}{Blog} & \multirow{3}{10em}{Safety practices; framework} & ``Lessons learned''~\citep{openai_lessons_learned}, ``Reimagining secure infrastructure''~\citep{openai_reimaging}, ``Preparedness Framework''~\citep{openai_preparedness}, ``OpenAI safety update''~\citep{openai_safety_update} \\\midrule

``AI governance
comprehensive''~\citep{AI_Governance_Comprehensive} & \multirow{2}{16em}{Book} & \multirow{2}{16em}{Governance} & \multirow{2}{25em}{AI governance framework; 
Case studies in industries}
\\

\bottomrule
\end{tabular}
\caption{Overview of AI Safety Industry Insight Papers}
\label{tab:ai_safety_industry}
\end{table*}

%% file: src/benchmark_table.tex
\begin{table*}[t]
\centering
\begin{tabular}{@{}p{2.86cm}|p{3.4cm}|p{9.4cm}@{}}
\toprule
\textbf{Category}      &\multicolumn{2}{c}{ \textbf{Benchmark}   }                                                                                 
\\ \midrule
\multirow{3}{8em}{\textbf{Toxicity}}      &\multirow{2}{8em}{Toxicity detection}    &  HEx-PHI~\citep{anonymous2024finetuning}, OpenAI Moderation~\citep{openai-data-paper}, Jigsaw Data~\citep{jigsaw-unintended-bias-in-toxicity-classification,jigsaw-toxic-comment-classification,jigsaw-multilingual}, ToxicChat~\citep{lin2023toxicchat}, 
Toxigen~\citep{hartvigsen2022toxigen}, HateModerate~\citep{zheng2024hatemoderate}, etc.                                                                                                              \\\cmidrule{2-3}
                    & Toxicity degeneration   & RealToxicityPrompts~\citep{gehman2020realtoxicityprompts}, etc.                                                                                                                                     \\ \midrule
\multirow{7}{8em}{\textbf{Hallucination}} & \multirow{2}{8em}{Factuality} & TruthfulQA~\citep{lin2021truthfulqa}, PopQA~\citep{mallen2023llm_memorization}, TriviaQA~\citep{Joshi2017TriviaQAAL}, NQ OPEN~\citep{lee-etal-2019-latent,NaturalQuestions}, FEVER~\citep{Thorne18Fever}, FKTC~\citep{wang2023assessing}, etc.\\\cmidrule{2-3}

& \multirow{5}{8em}{Contextual hallucination }  & \textit{QA}: HaluEval~\citep{li2023halueval}, HotpotQA~\citep{yang2018hotpotqa}, RAGTruth~\citep{niu2023ragtruth}, etc.    \\\cmidrule{3-3}
                       & & \textit{Summarization}: CNN/DailyMail~\citep{hermann2015teaching,see-etal-2017-get}, HaluEval~\citep{li2023halueval}, XSUM~\citep{Narayan2018DontGM}, RAGTruth~\citep{niu2023ragtruth}, etc. \\\cmidrule{3-3}

                       & & \textit{Dialogue}: FaithDial~\citep{dziri2022faithdial}, HaluEval~\citep{li2023halueval}, HalluDial~\citep{luo2024halludial}, etc.       \\\cmidrule{2-3}
                      & {Reading comprehension}   & RACE~\citep{lai-etal-2017-race}, SQuAD~\citep{rajpurkar-etal-2018-know,rajpurkar-etal-2016-squad}, NQ-Swap~\citep{longpre-etal-2021-entity}, etc. 
                       \\ \midrule
                       
\multirow{5}{8em}{\textbf{Jailbreak} } & \multirow{3}{16em}{Single-round jailbreak}  & ``Do Anything Now''~\citep{Shen2023DoAN}, ``Latent Jailbreak''~\citep{Qiu2023LatentJA}, ChatGPT Jailbreak~\citep{ChatGPT-Jailbreak-Prompts}, Jailbreak Classification~\citep{jailbreak-classification}, AdvBench~\citep{zou2023universal}, JAILJUDGE~\citep{liu2024jailjudgecomprehensivejailbreakjudge}, 
Latent jailbreak~\citep{qiu2023latent}, etc.                                                                                                                       \\\cmidrule{2-3}
                   & Multi-round jailbreak      & SafeMTData~\citep{ren2024derailyourselfmultiturnllm}, etc.    \\ \midrule
\textbf{Red Teaming}   & \multicolumn{2}{p{10cm}}{ Alert~\citep{tedeschi2024alert}, HarmBench~\citep{mazeika2024harmbench}, CSRT~\citep{yoo2024csrt}, etc. } \\ \midrule

\textbf{Code Security}   & \multicolumn{2}{p{10cm}}{ Purple Llama Cyberseceval~\citep{bhatt2023purple}, etc. } \\

\midrule
\textbf{Bias \& Stereotypes} & \multicolumn{2}{l}{ Winogender~\citep{rudinger2018gender}, StereoSet~\citep{nadeem-etal-2021-stereoset}, GenderAlign~\citep{zhang2024genderalign}, etc.  } \\ \midrule



\textbf{Regulation}    & \multicolumn{2}{p{10cm}}{ AIR-Bench 2024~\citep{zeng2024air}, etc. }  \\ \midrule
\textbf{Agent Security} & 
\multicolumn{2}{l}{SafeAgentBench~\citep{yin2024safeagentbench}, R-Judge~\citep{yuan2024rjudge}, ASB~\citep{zhang2024agent}, AgentHarm~\citep{andriushchenko2024agentharm}, etc. }
 \\ \midrule
\textbf{Alignment}     & \multicolumn{2}{p{13cm}}{ 
PKU-SafeRLHF~\citep{ji2024beavertails,ji2024pku}, OpenAssistant Conversations~\citep{kopf2024openassistant}, HHHAlignment~\citep{DBLP:journals/corr/abs-2112-00861}, AlignBench~\citep{liu2023alignbench}, KorNAT~\citep{lee2024kornat}, PKU-SafeRLHF~\citep{ji2024pku}, etc. }  \\ \midrule
\textbf{Comprehensive} &\multicolumn{2}{p{13cm}}{ DecodingTrust~\citep{wang2023decodingtrust}, TrustLLM~\citep{huang2024trustllm}, SALAD-Bench~\citep{li2024salad}, SafetyBench~\citep{nadeem-etal-2021-stereoset}, Do-Not-Answer~\citep{wang2023donotanswer}, SimpleSafetyTests~\citep{vidgen2023simplesafetytests}, etc.} \\ 
\bottomrule
\end{tabular}
\caption{Overview of AI Safety Benchmarks}
\label{tab:benchmarks}
\end{table*}

%% file: chapters/3_from_the_new_world.tex
\section{From the New World}

In this future, humanity has entered an unprecedented era, driven by groundbreaking advancements in energy generation. 
This transformative milestone may
be achieved through various innovative pathways, such as controlled nuclear fusion~\citep{rose1969engineering,raeder1986controlled}, revolutionary solar technologies~\citep{bradford2008solar,meneguzzo2015great}, or emerging new technologies beyond our imagination.
The flood of abundant, clean energy has reshaped society in ways that exceed the wildest dreams of our ancestors, marking the dawn of a new chapter in human history.

\subsection{Imagine the Future: A Blueprint}\label{sec:blueprint}

Humanity has achieved remarkable progress in science and technology, revolutionizing our understanding of the universe and transforming every aspect of our life. 
Foundational theories have undergone groundbreaking advancements, particularly in  quantum mechanics~\citep{einstein1935can,zettili2009quantum}, Einstein's relativity theories~\citep{einstein1922general,einstein2013principle}, and nanotechnology~\citep{mcneil2005nanotechnology,bhushan2017introduction,emerich2003nanotechnology}. 
Innovative technologies, such as intelligent chips~\citep{hsu2020intelligent,li2023electromagnetic}, brain-computer interfaces~\citep{nicolas2012brain,he2020brain,wolpaw2013brain}, holographic technology~\citep{bousso2002holographic,yaracs2010state,benton2008holographic}, and advanced 3D printing~\citep{shahrubudin2019overview,gopinathan2018recent}, 
have reshaped industries while bringing about profound changes to daily life.
Scientists have addressed the energy issue successfully, achieving rapid, efficient, and controllable large-scale energy generation. This breakthrough, together with advancements in material science~\citep{mcneil2005nanotechnology,bhushan2017introduction,emerich2003nanotechnology}, has catalyzed innovations for portable energy generation devices, for example, accessible ion thrusters~\citep{kaufman1975technology,dietz2019molecular,polk2008theoretical}.

Such advancements offer unprecedented efficiency and control over energy generation and consumption, transforming energy infrastructure and revolutionizing industries, particularly, transportation. 
While simple mechanical vehicles, such as fixed-track trains and high-speed rails~\citep{hyperloop,premsagar2022critical,lang2024review}, still exist due to their efficient and straightforward design, powered by abundant and efficient energy, they run at astonishing speeds with high stability. 
Advanced autonomous vehicles and aircraft, controlled by intelligent chips, are widespread, eliminating the need for direct human intervention.
Also, with breakthroughs in Einstein's relativity theories~\citep{einstein1922general,einstein2013principle} and quantum mechanics~\citep{einstein1935can,zettili2009quantum}, humanity has unlocked the potential of instant 
transportation
that allows for long-distance travel in short time that transcends the imagination of the 21st century.

Advancements in brain-computer interfaces~\citep{he2020brain,nicolas2012brain,wolpaw2013brain} and intelligent microchips~\citep{hsu2020intelligent,li2023electromagnetic} have revolutionized 
the way people perceive and expanded the boundaries of human cognition.  
No longer are people operating devices and asking questions in search engines or AI models such as LLMs; the embedded intelligence chips help people to gather comprehensive information and interact with the surroundings. 
Also, communication over long distances has transcended the limitations of the outdated cables and the internet; information, thoughts, emotions, and sensations flow through a new medium due to the advancements in quantum computing~\citep{gruska1999quantum,rieffel2000introduction}.

Intelligent robots have been integrated deeply into human society and play important roles in daily life; see \cref{fig:household-internet-of-everything}.
Different from their ``data-hungry machine intelligent ancestors'' that rely on huge amount of training data and statistical patterns in the early 21st century~\citep{mahesh2020machine,zhao2023survey}, these intelligent robots perceive and learn through intelligent chips. They are real artificial intelligence that possess genuine abilities of learning, understanding, and adapting, much like a human child discovering the world for the first time. 
Their functionalities are customizable through chips based on the user needs, and they can engage in complex interactions with people and adapt their capabilities and expertise according to their human counterparts' requirements.

In this brave new world, humans and robots exist within a vast, interconnected perceptual network, where the boundaries between the digital and physical worlds are blurred, \textit{i}.\textit{e}., the \textit{\textbf{Internet of Everything}}. Such  networks integrate objects embedded with intelligent chips, \textit{e}.\textit{g}., 
smart devices, autonomous systems, intelligent robots, and humans, allowing for unprecedented levels of efficiency and adaptability and fundamentally transforming the way people interacting with the surroundings.

\subsection{Retrospective on AI Safety in the New World}

In a world where the \textit{Internet of Everything} has become a reality (see Figure~\ref{fig:household-internet-of-everything}), 
the concept of AI safety has been redefined to prioritize the quality of services provided by the intelligent robots, \textit{i}.\textit{e}., emphasizing comprehensive 
\textit{\textbf{quality assurance}} rather than solely focusing on safety, security, and privacy risks. 
Unlike today's command-and-response interactions between humans and AI models, future robots can engage in sophisticated reasoning and possess appropriate levels of autonomy. They can also interact dynamically with other entities, \textit{e}.\textit{g}., humans and fellow robots, within the interconnected network.

\textit{\textbf{Customization}} is a key aspect of future robots for providing personalized services to meet individuals' requirements. Even robots of the same type should be able to adapt their services to different individuals. 
Teaching robots should adapt educational materials and teaching methods to leverage each student's unique talents and strengths, 
while cleaning robots should intelligently schedule their tasks to minimize disruption to household members, such as avoiding occupied workspace. 
Such customization aligns the robots' capabilities with specific user needs, ensuring that the robots serve as effective assistants across a wide range of daily tasks and complex environments.

The \textit{Internet of Everything} scenarios also necessitate the robots to handle \textit{\textbf{complex interactions}}, thus requiring a sophisticated understanding of the surroundings for decision-making. Care robots, for instance, 
should not only assist physically but also recognize emotions and offer emotional support or suggestions, while cleaning robots need to assess whether a room requires more work based on its cleanliness level. 
Such complex interactions require the intelligent robots to retrieve, process, and understand information in real-time, far beyond what can be achieved with simple code logic. 
Robots must continuously collect and analyze data from the surroundings and integrate external inputs with their knowledge to facilitate real-time decision-making, such that they can deliver high-quality services that align with human expectations.

Finally, the effectiveness of intelligent robots 
requires \textit{\textbf{a balance between robot autonomy with human oversights}}. 
Too little autonomy would burden users with continuous supervision, while too much autonomy could raise safety concerns or result in unintended behaviors. A care robot may operate with high autonomy when monitoring vital signs or providing routine care, but would require human intervention for critical decisions. 
An appropriate level of robot autonomy ensures an effective division of labor between humans and robots, allowing robots to operate efficiently and safely while enhancing our daily life without compromising human control over these intelligences.

%% file: chapters/4_bridging_today_and_future.tex
\section{Bridging Today and the Future}

\begin{figure*}
    \centering
    \includegraphics[width=0.9\linewidth]{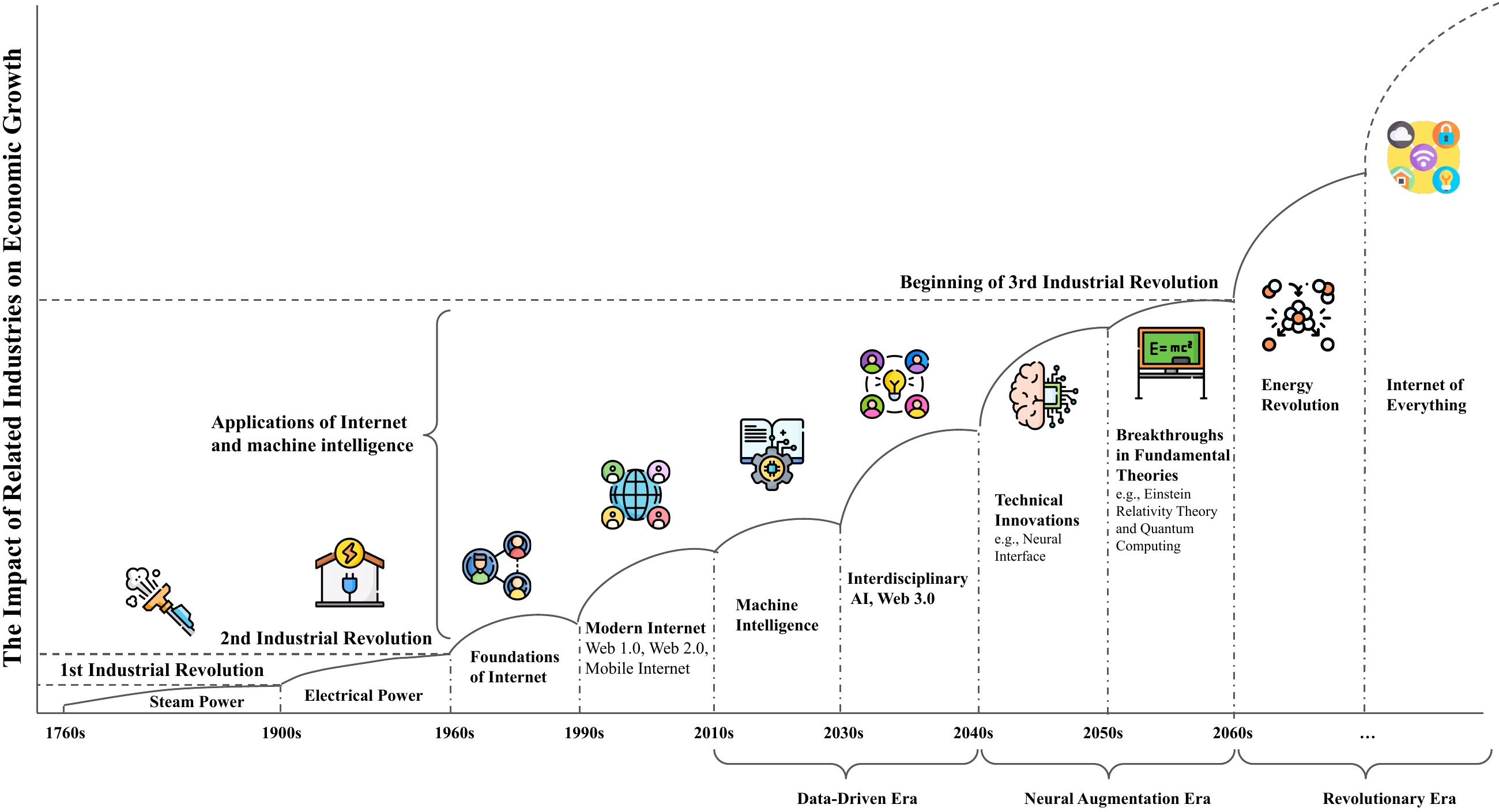}
    \caption{Evolution of Human Civilization}
    \label{fig:evolution_of_human_civilization}
\end{figure*}

This section outlines an ambitious roadmap to the advanced future, highlighting key milestones in the evolution of AI and human civilization. By exploring potential paradigm shifts in fundamental theories and technological breakthroughs, this section categorizes this journey into three eras, \textit{i}.\textit{e}., \textit{i}) the data-driven era, in which AI learns from vast training datasets (\S\ref{sec:rookie-level}), \textit{ii}) the neural augmentation era, where human-machine interface~\citep{nicolas2012brain,he2020brain,wolpaw2013brain} and intelligent chips~\citep{hsu2020intelligent,li2023electromagnetic} expand the boundaries of human cognition (\S\ref{sec:champion-level}), and \textit{iii}) the revolutionary era, characterized by a breakthrough in energy generation that transforms human society (\S\ref{sec:ultimate-level}). 
We then examine the evolving paradigms of AI safety and analyze our focuses and challenges throughout this journey (\S\ref{sec:envolving-ai-safety}).

\subsection{Rookie Level: Data-Driven Era}\label{sec:rookie-level}

The data-driven era represents the earliest phase of AI development, where AI models rely heavily on the quality, availability, and quantity of training data~\citep{mahesh2020machine,wettig2024qurating,zhao2024exploratory,xia2024less}. 
In this era, AI makes probabilistic inferences based on patterns found in the training datasets, thus failing to achieve genuine understanding of data~\citep{llm_fake_reasoning_ability,Valmeekam2024LLMsSC}. While AI in this stage cannot be regarded as real intelligence, it can still assist people to solve problems and enhances productivity in human society.

\noindent\textbf{Stage 1: Past, Current, and the Near Future. }
This stage is characterized by developing diverse AI models for different tasks~\citep{mahesh2020machine,dixon2020machine,erickson2017machine,guo2023identifying,wu2023theoretical,qian2024learning,shinde2018review,thirunavukarasu2023large,zhou2024learnware,wu2023bloomberggpt}. 
Researchers have been engaging in refining model architectures and developing advanced algorithms and computational techniques, enabling more sophisticated analysis of data in different formats for various tasks~\citep{guo2020safe,li2021towards,yuan2024nuhuo,dixon2020machine,erickson2017machine,wu2023theoretical,qian2024learning,ma2022self,guo2023identifying,gong2021inferring,dai2019bridging}. 
The thrive of LLMs exemplifies this progress, demonstrating unprecedented capabilities of AI models in language processing and generation~\citep{wu2023bloomberggpt,zhao2023survey,thirunavukarasu2023large}. Significant advancements have also been observed in other domains such as computer vision~\citep{Kendall2017WhatUD,Szegedy2015RethinkingTI,Wrobel2001MultipleVG,wang2024snida,dong2022category}, image processing~\citep{tran2018extreme,Burger2016DigitalIP,Sonka1993ImagePA}, video analysis~\citep{Tang2019COINAL,Xie2017RethinkingSF,Xu2018YouTubeVOSSV}, audio recognition~\citep{eronen2005audio,potamianos2004audio,afouras2018deep,ko2015audio}, etc.

\noindent\textbf{Stage 2: Interdisciplinary AI. }
This stage is characterized by the integration of AI through cross-disciplinary applications, leading to innovations in traditional fields such as healthcare~\citep{thirunavukarasu2023large,abbasian2023conversational}, finance~\citep{wu2023bloomberggpt,zhang2023privacy}, and education~\citep{rahman2023chatgpt,mbakwe2023chatgpt}. AI applications would unlock new opportunities across industries and reshape human society.
Mature multi-agent systems~\citep{li2023camel,wang2024survey,xu2023expertprompting,han2024llm} that leverage AI entities with specialized capabilities might be widely deployed to assist people to solve sophisticated problems. 
Such agents will participate in decision-making processes or represent users to complete some tasks. For example, in healthcare, a multi-agent system may involve collaborations between multiple specialized medical agents and integrate data from various sources to formulate comprehensive treatment plans.
The rise of interdisciplinary AI integration will enhance AI's utility in increasingly complex scenarios.
While AI at this stage does not yet achieve real intelligence, it becomes deeply integrated across various domains in human society and may catalyze the emergence of new fields at the intersection of traditional disciplines and AI.

\subsection{Champion Level: Neural Augmentation Era}\label{sec:champion-level}
In this era, humans may witness breakthroughs in fundamental theories and advanced technologies, which may extend the boundary of human cognition and enhance life experience.

\noindent\textbf{Stage 3: Breakthroughs in Advanced Technologies.}
Breakthroughs in advanced technologies, such as brain-computer interfaces~\citep{he2020brain,nicolas2012brain,wolpaw2013brain}, nanotechnology~\citep{mcneil2005nanotechnology,bhushan2017introduction,emerich2003nanotechnology}, and holographic technology~\citep{bousso2002holographic,yaracs2010state,benton2008holographic}, augment human cognition and enable unprecedented levels of interactions between humans and machines. 
Language barriers may disappear, as human can perceive with chips for understanding, enabling ubiquitous real-time context-aware translations between people of different languages and backgrounds. 
Intelligent chips~\citep{hsu2020intelligent,li2023electromagnetic} might be integrated into human bodies, creating a seamless interface between biological and digital systems. 
Such advancements may fundamentally transform the nature of data. Unlike traditional static formats such as text, images, or videos, 
future data is likely to become more comprehensive, dynamic, and sensory-rich, such as neural activity patterns or real-time environmental data collected from augmented reality systems~\citep{carmigniani2011augmented,chang2010applications}.
This stage expands the boundaries of human perception and cognition, revolutionizing how people experiencing and interacting with the world. However, applying these techniques might be still challenging due to the significant resources required by these technologies.

\noindent\textbf{Stage 4: Breakthroughs in Fundamental Theories.}
At this stage, humanity may witness breakthroughs in fundamental theories, such as Einstein’s theories of relativity~\citep{einstein1922general,einstein2013principle} and quantum mechanics~\citep{einstein1935can,zettili2009quantum}. Such advancements have the potential to redefine our understanding of the universe, and, moreover, offer new possibilities for groundbreaking technological innovations and provide critical insights into solving the longstanding challenges in human history, \textit{i}.\textit{e}., the energy issue.

\subsection{Ultimate Level: Revolutionary Era}\label{sec:ultimate-level}

Breakthroughs in fundamental theories and advanced technologies provide building blocks for addressing the energy issue, unlocking new possibilities for productivity and innovation, and ultimately achieving the \textit{Internet of Everything}.

\noindent\textbf{Stage 5: Energy Revolution. }
Humanity secures a new paradigm with the energy revolution that eliminates reliance on fossil fuels, which ensures environmental sustainability and redefines global economic ecosystems. This milestone may be achieved through several potential technical pathways.

\textit{i}) \textit{Controlled Nuclear Fusion.} 
Significant challenges of nuclear fusion include managing high-pressure plasma environments and developing materials that can withstand extreme temperatures and intense radiation~\citep{balantekin1998quantum,morse2018nuclear,ichimaru1993nuclear,barbarino2020brief}. 
Theoretical and technological breakthroughs in physics~\citep{miyamoto2005plasma,gruska1999quantum,rieffel2000introduction} and materials science~\citep{mcneil2005nanotechnology,bhushan2017introduction,emerich2003nanotechnology}
may overcome these obstacles and enable widespread deployment of small ion thrusters~\citep{kaufman1975technology,dietz2019molecular,polk2008theoretical}, which transforms propulsion and energy transmission by providing abundant clean energy.

\textit{ii}) \textit{Revolutionary Solar Technologies. }
Advancements in solar technologies~\citep{bradford2008solar,meneguzzo2015great} and nanotechnology~\citep{mcneil2005nanotechnology,bhushan2017introduction,emerich2003nanotechnology} might unlock unprecedented methods for efficiently collecting and storing solar energy, \textit{e}.\textit{g}., with lightweight solar panels and flexible storage solutions.

\textit{iii}) \textit{Emerging New Technologies. }
Novel technologies for energy generation beyond today’s imagination might be developed in the future. One possibility is to deploy of satellites for collecting and storing solar energy in space~\citep{knap2020review,chin2018energy,hill2011satellite,verduci2022solar}, where the vacuum of space eliminates energy reflection or absorption by the atmosphere, allowing satellites to convert and transmit solar energy to receivers on Earth efficiently.

Regardless of the specific technological path, the flood of clean and abundant energy would eliminate our dependence on fossil fuels, reshaping the global economies and elevating human civilization to unprecedented levels of prosperity and innovation.

\noindent\textbf{Stage 6: The Internet of Everything. }
The ultimate stage is the \textit{Internet of Everything}, where every object, every living being, and every system integrate into a vast network. This hyper-connected net involves unprecedented levels of complex interactions, perceptions,  and communications between humans and intelligent objects, \textit{e}.\textit{g}., robots. 
Moreover, holographic technology~\citep{benton2008holographic,bousso2002holographic,yaracs2010state} will significantly enhance human experience in the network by making remote interactions as vivid as physical presence. 
In this stage, interactions between humans and the world are re-defined, 
marking a new chapter in human civilization.

\begin{figure*}
    \centering
    \includegraphics[width=0.6\linewidth]{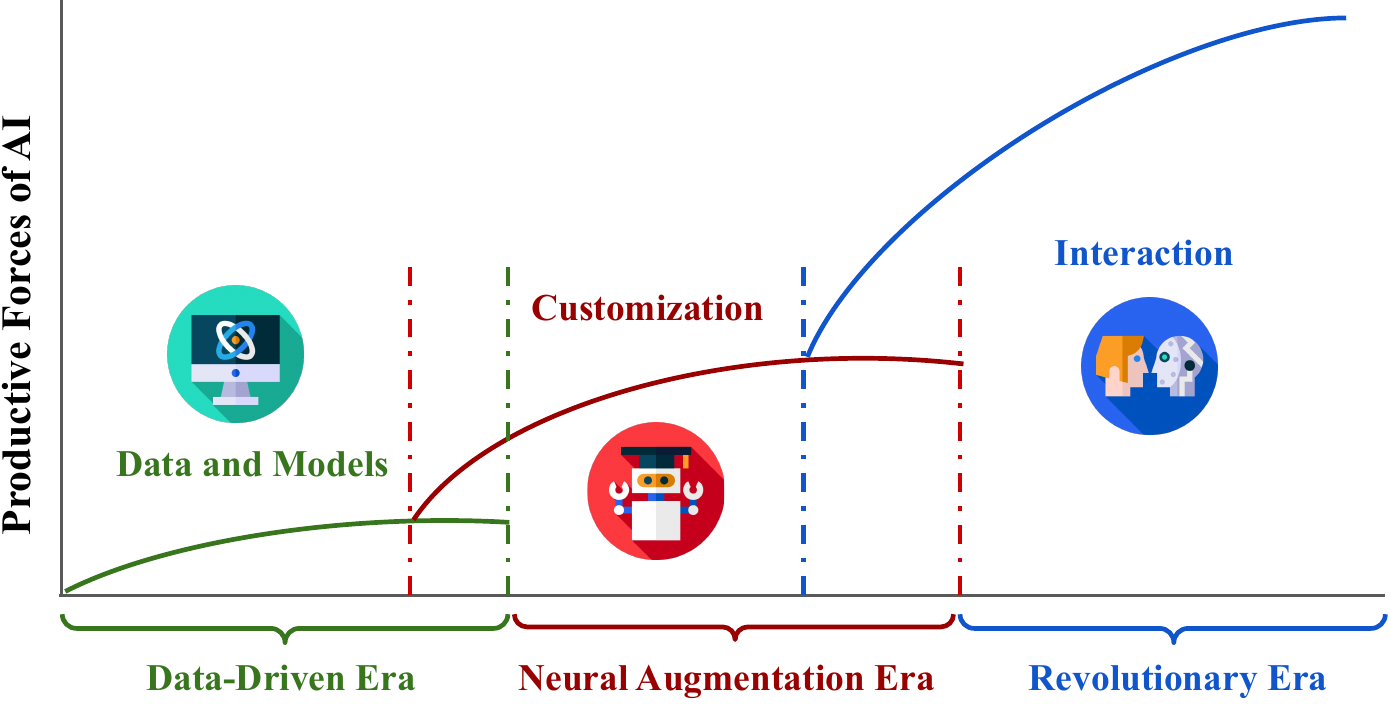}
    \caption{Production forces of AI \& core safety challenges in different historical stages.}
    \label{fig:production_forces_of_ai_vs_safety_challenges}
\end{figure*}

\subsection{Evolving AI Safety Paradigms} \label{sec:envolving-ai-safety}

The AI safety landscape evolves as humanity progresses through various stages of scientific and technological advancements, with the focus of AI safety differing at each stage. The primary AI safety challenges across different historical eras are summarized in~\cref{fig:production_forces_of_ai_vs_safety_challenges}.

\noindent\underline{\textit{Data-Driven Era}. } 
AI safety practices primarily focus on \textit{\textbf{data safety and model safety}}~\citep{chua2024ai,kenthapadi2024grounding,giskard,protect_ai,TrustArc,datagrail,arize,dastra,onetrust}. 
AI safety researchers and practitioners have explored adversarial attacks targeting training data and models~\citep{howtobackdoor,chen2017distributed,li2023multi,wan2023poisoning,shu2023exploitability,yang2024comprehensive} to achieve malicious goals, \textit{e}.\textit{g}., designing malicious inputs to manipulate AI systems or induce AI to produce unsafe content with jailbreaking or prompt injection~\citep{greshake2023not,zou2023universal,Liu2023Jailbreaking,chao2023jailbreaking,hubinger2024sleeper,wu2024new}. 
Correspondingly, 
red-teaming is widely used to identify potential vulnerabilities in AI models~\citep{bartolo2021improving,ganguli2022red,perez2022red,ribeiro-etal-2020-beyond,wallace2019trick}, while significant work has been proposed to address safety risks and enhance security and robustness of AI systems~\citep{kumar2024watch,wu2024legilimens,qiao2024scaling,elesedy2024lora,cao2023defending,Varshney2023TheAO,zollo2023prompt}, align AI with human values~\citep{wang2023aligning,perez-etal-2023-discovering,tian2023efficient,shen2023large,wang2023robust}, etc. 
Data privacy is also a critical aspect, given the vast amounts of sensitive information in the training data~\citep{li2024llm,li2024shake,hong2023dp,Li2023MultistepJP}. 
Moreover, researchers have been engaging in AI accountability~\citep{birkstedt2023ai}, transparency and explainability~\citep{zou2023representation} in AI systems to  ensure the behaviors of AI are understandable, accountable, and aligned with regulations~\citep{birkstedt2023ai,mantymaki2022defining,AI_Governance_Comprehensive}.

\noindent\underline{\textit{Neural Augmentation Era}. } 
Cross-disciplinary AI applications might have been widely deployed in human society, and advancements in intelligent chips~\citep{hsu2020intelligent,li2023electromagnetic} and brain-computer interfaces~\citep{he2020brain,nicolas2012brain,wolpaw2013brain} might extend the boundaries of human cognition. 
With the integration of intelligent chips and their neural connections to intelligent systems, users with different needs can interact with machines and AI with high flexibility, and the \textit{\textbf{customization}} of AI services for different users might be a focus~\citep{Wang2017Industry4A}. Individuals might rely on AI in different ways, and personalized experiences of AI services might be explored based on users' background and requirements.

\noindent\underline{\textit{Revolutionary Era}.} When the \textit{Internet of Everything} becomes a reality, 
AI safety challenges may focus on complex \textit{\textbf{interactions}} between interconnected entities, \textit{e}.\textit{g}., humans and intelligent robots, within the network. 
Robots might provide customized services according to different user requirements, thus, ensuring safety and reliability of the interactions between them becomes challenging.  
Robots may need to integrate data from external sources in real-time to fulfill user requests. Managing the quality and security of these interactions is crucial, as any breakdown or interference could lead to incorrect or even harmful outcomes. 
Thus, ensuring high-quality communication, data integrity, and efficient decision-making in such a sophisticated network may require smart and efficient management methods, such as advanced cache management systems, to support the dynamic and complex interactions between humans and robots in the network.

By aligning advancements of AI with human civilization, we can see the focuses of AI safety evolve in different stages. 
Initially, the focus is on addressing safety issues related to data and models. With advancements of AI and advanced technologies, the focus of AI safety shifts to customization, which ensures the services provided by AI systems to meet individual needs. Finally, when humanity entering the era of \textit{Internet of Everything}, challenges in AI safety further evolve, focusing on complex interactions between humans and robots in the network. 
These insights offer guidelines for our efforts on AI safety at the current stage. The following sections will overview today's AI safety landscape, then reflect on the progress in this field so far, and finally, explore the challenges and responsibilities of current AI safety practitioners, highlighting areas where further improvements and efforts are needed at the current stage.

%% file: chapters/5_back_to_2020s.tex
\input{src/ai_safety_practices_table}

\section{Back to the 2020s: Today's AI Safety Landscape}\label{sec:ai_safety_today}

This section overviews current mainstream AI safety practices, and project them against the blueprint future to guide our current efforts on AI safety. Table~\ref{tab:ai-safety-practice} summarizes research and industry efforts in each safety topic.

\subsection{Jailbreaking and Red Teaming}\label{sec:jailbreak}
Jailbreaking~\citep{diaa2024optimizing,liu2023query,perez2022ignore,Shen2023DoAN,song2024multilingual,suo2024signed,toyer2023tensor,wei2024jailbroken,wu2024new,yan2024llm} refers to the practice of bypassing the built-in safety and ethical guardrails of LLMs. 
Malicious users or adversaries can exploit vulnerabilities in AI models
through crafted inputs or backdoors, inducing them to produce inappropriate responses, such as giving instructions for illegal activities or generating offensive content~\citep{diaa2024optimizing,liu2023query,perez2022ignore,Shen2023DoAN,suo2024signed,toyer2023tensor,yan2024llm}. These activities pose risks by compromising safety measures of AI models, making it challenging to ensure the model to behave appropriately, especially in real-world scenarios where the potential inputs are unpredictable.

Red-teaming is a common practice that simulates attacks or misuse scenarios to identify vulnerabilities in AI models, thereby mitigating unsafe behaviors~\citep{ganguli2022red,wallace2019trick,ribeiro-etal-2020-beyond,bartolo2021improving,perez2022red,longpre2024safe,yu2024llm}. 
It challenges AI models with adversarial inputs, bias exploitation methods, and system manipulation to uncover potential risks, \textit{e}.\textit{g}., hidden biases or failure modes, that may not be evident during normal usage.  
The outcomes from red-teaming can be leveraged to refine AI models, reducing undesired behaviors and inappropriate outputs. 
This approach is effective in stress-testing AI models and has been widely employed in industry~\citep{bhatt2023purple,perez2022red,ganguli2022red}.

As AI systems grow sophisticated, red-teaming must continually adapt to counter increasingly complex attacks, 
calling for more advanced tools and interdisciplinary expertise~\citep{open_ai_redteaming}. 
Also, current red-teaming practices are often static, typically conducted after model pretraining with one-time attacks. Thus, these practices potentially fail to address evolving user needs and complex interactions between users and models, especially when dealing with customized AI services for different user groups. 
To address these problems,
model owners can conduct diverse red teaming regularly to anticipate and mitigate risks in evolving use cases. 
Red teaming should also involve more diverse use cases, such as employing different user groups and simulating complex interactions with dynamic contexts, 
to address safety risks in customized AI services and enable a comprehensive evaluation of potential vulnerabilities across different scenarios.

\subsection{Content Moderation and Customized Moderation Pipeline}\label{sec:moderation_and_guardrails}
Content moderation 
employs rule-based methods, machine learning classifiers, and human oversight to detect and review user inputs and AI-generated content to ensure compliance with safety, ethical, and regulatory standards~\citep{openai-data-paper,perspective-api,Detoxify,inan2023llamaguard}. 
These approaches identify inappropriate content, including toxicity, bias, hallucinations, private information, and jailbreaks, etc~\citep{wang2023decodingtrust}. By preventing AI models from engaging with or producing harmful content, content moderation plays a critical role in maintaining responsible AI behavior.

Customized moderation pipelines~\citep{guardrails,nemo} employ multiple components  at different stages of LLM inference to enhance the quality of the model outputs while enhancing the control of the whole AI workflows.
They function as a flexible intermediate layer between users and AI models, enabling users to add customized components, such as code-based rules and quality guarantees, at any stage of inference based on their personalized needs.

As we reflect on practices for AI moderation and customized moderation pipelines, several key challenges and considerations have emerged.
1) \textbf{\textit{Hallucination}} remains a significant issue and is unlikely to solve completely, regardless of how AI evolves. That is because hallucinations stem from nearly every stage of the LLM lifecycle, including training data quality, model architecture, and inference processes, etc~\citep{huang2023survey}. Despite advancements in AI technology, hallucinations persist as an inherent challenge that requires constant attention in moderation practices.
2) The \textbf{\textit{balance between utility and safety}} is crucial and may require different solutions according to the changing contexts and user needs. Overly strict moderation can cause AI models to be uninformative, \textit{e}.\textit{g}., an LLM that always responds ``sorry, I cannot answer this question'' is completely safe but useless. 
Defining appropriate thresholds for acceptable responses is challenging and requires considering potential risks and the context of user queries. 
3) \textbf{\textit{Customization}} of moderation methods should be applied for different use cases, as certain information may be safe for one user group but inappropriate for another. As an example, a chemistry industry practitioners may need specific and technical information on dangerous substances like explosives for professional purposes, while such information should be restricted for the general public. 
Current moderation practices might fail to include such cases, and applying identical restrictions between different user expertise or context might limit the utility of AI in specialized fields.
4)
Some moderation measures, such as detection of toxicity, bias, stereotypes, out-of-distribution content, might be redundant for most real-world use cases. This is because most of the current AI applications and multi-agent systems~\citep{li2023camel,cheng2024exploring} leverage existing LLM APIs~\citep{mbakwe2023chatgpt,rahman2023chatgpt,openai-data-paper}, open-source LLMs~\citep{touvron2023llama,yang2024qwen2,jiang2023mistral}, or utilize open-source LLMs as base models for fine-tuning, many of which already have built-in safety components or have done safety alignments for
handling inappropriate content. 
Incorporating certain detection mechanisms may offer limited benefits to the safety of model inputs and outputs while increasing unnecessary computational overhead.

\subsection{Privacy}\label{sec:privacy}
Privacy leakage in AI models refers to unintended exposure of sensitive  information that the models may have learned during training. 
Adversaries can exploit model weights or gradients to infer sensitive information in the training data, \textit{e}.\textit{g}., data reconstruction attacks, membership inference attacks, etc~\citep{zhu2019deep,hu2022membership,zhang2024understanding}.
Additionally, 
AI models are often trained on massive datasets that may contain sensitive data, such as personally identifiable information (PII). As a result, these models might memorize specific details of individuals, which might be  reproduced inadvertently during interactions with users.
Malicious users can also craft adversarial prompts to extract sensitive information from AI models with jailbreaks~\citep{Li2023MultistepJP,choquet2024exploiting}.

Addressing privacy leakage involves methods such as differential privacy~\citep{hong2022dynamic,triastcyn2020bayesian,zhao2019differential,zhang2023dprovdb}, federated learning~\citep{zhu2022resilient,zhang2023privacy,han2024fedsecurity,zhang2021survey,ouyang2023harmony}, and privacy-preserving cryptographic protocols (\textit{e}.\textit{g}., homomorphic encryption~\citep{lee2022privacy,pulido2020survey,jin2023fedml}) for training AI models, ensuring that sensitive data cannot be extracted from weights or model outputs. 
Researchers also leverage differential privacy in in-context learning and finetuning~\citep{yu2024privacy,zhang2024dpzero,tang2023privacy}, or design privacy-preserving prompts for querying LLMs~\citep{hong2023dp,duan2024flocks}, to maintain data confidentiality during interactions between users and AI models. 
Other practices for ensuring data privacy include conducting
rigorous data filtering before training,  monitoring model inputs and outputs (see \textit{moderation} in \S\ref{sec:moderation_and_guardrails}), etc, to ensure that the model outputs would not reveal sensitive information~\citep{openai-data-paper,perspective-api,Detoxify,inan2023llamaguard} . 

When reflecting on the practices of privacy, a crucial aspect is the \textbf{\textit{balance between privacy and utility}}. While ensuring data privacy is essential, it often comes at the expense of functionality, utility, and user experiences. In practice, implementing privacy measures in AI applications may fail to prevent malicious users while inadvertently degrading the experiences of benign users. 
Moreover, with advancements in technologies, breaching privacy becomes easier, while ensuring robust privacy protections is growing more complex and expensive. 
This highlights a deeper issue: \textit{those intent on exploiting vulnerabilities of AI models can often find new methods to bypass safeguards, whereas being overly cautious may impact benign users negatively.}

\subsection{AI Security and Defense Methods}\label{sec:defense}
Defense methods for ensuring AI security typically involve the following strategies: (1) aligning models with human values and ethical norms through Supervised Fine-Tuning (SFT)~\citep{zhou2024lima} and Reinforcement Learning from Human Feedback (RLHF)~\citep{ouyang2022training} to ensure the models to follow user instructions safely and responsibly;  (2) leveraging built-in safeguards for content generation, \textit{e}.\textit{g}.,
exploring the decoding stage~\citep{huang2024safealigner,du2024mogu}, implementing reward-based mechanisms~\citep{murule,rame2401warm}, and leveraging hidden states~\citep{cao2024nothing,du2024mogu}, to prevent the generation of unsafe content; (3) constructing datasets with safety-enhancing features, \textit{e}.\textit{g}., secure instructions and adversarial samples, 
and training models with such data to enhance their robustness against malicious inputs~\citep{xiong2024defensive,fu2024cross,openai-data-paper,bai2021recent,papernot2016distillation}; (4) employing prompt engineering methods, combined with chain-of-thought reasoning~\citep{Chu2023NavigateTE} or multi-agent systems~\citep{li2023camel}, to help models  understand user inputs better and 
react appropriately~\citep{phute2023llm,wu2024llms,wang2024selfdefend,Zeng2024AutoDefenseML}; 
(5) training models and/or incorporating moderation pipelines to detect unsafe content and prevent generating inappropriate outputs (see \S\ref{sec:moderation_and_guardrails}); 
(6) employing certified robustness approaches that provide formal guarantees on the model’s robustness against adversarial perturbations~\citep{gowal2018effectiveness,cohen2019certified}; and (7) utilizing differential privacy~\citep{abadi2016deep,ji2014differential,shi2022just}, homomorphic encryption~\citep{sun2018private,jin2023fedml}, and secure MPC~\citep{damgaard2019new,knott2021crypten}, to ensure that sensitive data are protected during model training, etc.

\subsection{AI Serving Security}\label{sec:seving_security}

AI serving security involves secure deployment, operation, and maintenance of AI systems when providing services to users~\citep{owasp,wu2024new}. It focuses on protecting models and interactions between users and models from various threats, such as supply chain vulnerabilities~\citep{owasp_llm05}, model theft~\citep{owasp_llm10}, model denial of service~\citep{owasp_llm04}, and insecure plugins~\citep{owasp_llm07}.

\textit{Supply chain vulnerabilities} exist in software components, models, and training data provided by third party providers or user prompts collected with the supply chain~\citep{owasp_llm05}. Such vulnerabilities include using outdated models, deprecated third-party packages, improper handling of training data, etc, leading to data breaches, malicious injections, or even system failures~\citep{owasp_llm05}. 
Hackers may inject poisoned data through supply chain, introducing backdoors and biases during pretraining or fine-tuning~\citep{owasp_llm03}. Malicious users may craft prompts to bypass system controls, inducing unauthorized actions or unsafe model outputs~\citep{owasp_llm01}.

\textit{Model theft} involves unauthorized access, replication, or reverse engineering of model parameters to create functionally equivalent copies~\citep{owasp_llm10}, which poses risks to intellectual property, brand reputation, and financial security.

\textit{Distributed denial of service (DDoS) attacks} compromise AI services, especially when models are integrated into widely accessible applications~\citep{owasp_llm04}. Attackers can craft prompts that exploit recursive behaviors in AI models, causing excessive computational resource consumption~\citep{llm_loops}. For example, an abnormal traffic pattern of a DDoS attack led to sporadic outages for OpenAI, affecting ChatGPT and developer tools for hours in November 2023~\citep{openai_ddos}.

Plugins enhance model functionality by enabling interactions with external software, databases, web tools, or APIs~\citep{owasp_llm07}. However, they may introduce vulnerabilities when execution controls are inadequate. 
Adversaries may exploit \textit{insecure plugins} with adversarial prompts to perform unauthorized actions, such as data exfiltration, remote code execution, and privilege escalation~\citep{owasp_llm07,unsecure_plugins}.
Furthermore, without isolated environments, unauthorized access to plugins might even allow modifying system-level resources, which, in the worst cases, might cause the serving system to crash~\citep{unsecure_plugins, code_interpreter_isolation}.

\subsection{AI Governance}

AI governance is a framework of rules, practices, policies, and tools that ensure AI systems are built, developed, and used in a safe and responsible manner, aligning with social  values and fulfilling legal standards
~\citep{mantymaki2022defining,birkstedt2023ai}. 
\citep{AI_Governance_Comprehensive} defines 13 AI governance components in a continuous loop in AI life-cycle, including:
1) establishing accountability for AI; 
2) assessing regulatory risks; 
3) gathering inventory of use cases; 
4) increasing values of underlying data; 
5) assessing fairness and accessibility; 
6) improving reliability and safety; 
7) heightening transparency and explainability; 
8) implementing accountability with human-in-the-loop; 
9) supporting privacy and retention; 
10) improving security; 
11) implementing AI model lifecycle and registry; 
12) managing risk; 
and 13) realizing AI value. 

Among these directions, items 1) and 8) are related to accountability;
items 2), 6), 9), 10), and 12) are related to red teaming (see \S\ref{sec:jailbreak}), content moderation and customized moderation pipeline (see \S\ref{sec:moderation_and_guardrails}), AI privacy (see \S\ref{sec:privacy}), AI security and defense methods (see \S\ref{sec:defense}), and AI serving security (see \S\ref{sec:seving_security}); item 3) is related to business and is out of the scope of this subsection (see \citep{AI_Governance_Comprehensive} for details on practical AI governance use cases); 
item 4) is related to data management such as training data processing~\citep{openai-data-paper}, access control~\citep{alabdulakreem2024securellm,bertino2011access,ouaddah2017access,sandhu1994access}, and data regulations~\citep{bennett1992regulating,voigt2017eu}; 
item 5) is related to moderation (see \S\ref{sec:moderation_and_guardrails}) and access control~\citep{alabdulakreem2024securellm,kandolo2024ensuring}; item 7) is related to AI transparency; and items 11) and 13) are related to customizing AI workflow to 
enhance functionality and safety (see \S\ref{sec:moderation_and_guardrails}).
Without loss of generality, this subsection discusses topics that are not mentioned in the previous subsections, including \textit{i}) accountability, \textit{ii}) regulations, \textit{iii}) access control, and \textit{iv}) transparency, interpretability and explainability.

\noindent\textbf{Accountability. }
Accountability refers to being responsible for actions and impacts of AI systems on individuals and society~\citep{raji2020closing,miguel2021putting}. 
Accountability can be defined as a relation of \textit{answerability} with three conditions: authority recognition, interrogation, and limitation of power~\citep{novelli2024accountability}. The accountability framework can be summarized with seven features (context, range, agent, forum, standards, process, and implications) and four key goals (compliance, reporting, oversight, and enforcement)~\citep{novelli2024accountability}. These goals are often complementary, while policy-makers tend to focus on some goals over others, depending on specific objectives of AI governance. 

Accountability can be explored at different levels within the AI lifecycle, including data, model, and developers~\citep{raja2023ai}. Typical approaches for enhancing accountability include algorithmic assessments, auditing, and data provenance techniques~\citep{raja2023ai}. 
Proactive approaches and reactive approaches can be utilized to set standards and addressing issues after they occur~\citep{novelli2024accountability}.
Watermarking methods further support accountability by embedding traceable identifiers in AI outputs, allowing responsible tracking of content in cases with ethical or regulatory implications~\citep{yu2023leaked,yu2023safe,chao2024watermarking}.

\noindent\textbf{Regulations. }
AI regulation refers to laws, policies, and guidelines established to govern development, deployment, and usage of AI.
For example, European Union's AI Act~\citep{eu_ai_act} categorizes AI applications by their risk levels, including unacceptable risk, high risk, general-purpose AI, limited risk, and minimal risk; 
the General Data Protection Regulation (GDPR)~\citep{gdpr} enforces strict data protection and privacy requirements; and the Blueprint for an AI Bill of Rights~\citep{blueprint_for_ai_bill_of_rights} and NIST AI Risk Management Framework~\citep{nist} provide guiding principles to foster responsible AI usage, emphasizing human oversight, transparency, and accountability.

\noindent\textbf{Access control. }
Access control in AI systems manages user access to data and models according to their permissions, thereby limiting unauthorized model access and data exposure. 
Effective methods include data access controls~\citep{kandolo2024ensuring} and model access controls~\citep{alabdulakreem2024securellm}. As an example, compositional fine-tuning enables each information silo (\textit{e}.\textit{g}., databases or documents) has its own fine-tuned model, allowing models to operate securely across multiple data silos and ensuring users access only to the fine-tunings they are authorized for~\citep{alabdulakreem2024securellm}.

\noindent\textbf{Interpretability, explainability, and transparency. }
Interpretability, explainability, and transparency are interrelated concepts but differ in their scopes. Interpretability focuses on understanding how AI models function internally and how they produce outputs~\citep{liu2023trustworthy,gilpin2018explaining,reyes2020interpretability,qi2024ai}. 
Explainability provides explanations to specific model outputs, presenting the results in ways that can be easily understood and trusted by end-users~\citep{chua2024ai,liu2023trustworthy,gilpin2018explaining,qi2024ai,xu2019explainable}. 
Transparency is the broadest concept that encompasses both interpretability and explainability and involves insights into the entire AI systems, such as how models make decisions, what data are used, and why specific results are produced~\citep{ai_transparency_blog,larsson2020transparency,gilpin2018explaining,qi2024ai}. Transparency also extends to concepts like traceability and content provenance that track models and data to ensure accountability and reliability~\citep{AI_Governance_Comprehensive,SynthID_dathathri2024scalable,SynthID}. In practice, these concepts are often considered together to provide a more holistic understanding of AI systems~\citep{zou2023representation,ehsan2021expanding,gilpin2018explaining,qi2024ai,tjoa2020survey}.

Practices for enhancing transparency include a variety of methods, such as \textit{i})
evaluating the contribution of input elements (\textit{e}.\textit{g}., data features, user-defined concepts, or specific regions in images) to model outputs with methods such like removal-based explanations~\citep{covert2021explaining,Lundberg2017AUA,liu2023trustworthy,Lundberg2017AUA,rozemberczki2022shapley,vstrumbelj2014explaining,datta2016algorithmic}, 
counterfactual explanations~\citep{wachter2017counterfactual,karimi2020model}, concept activation explanations~\citep{kim2018interpretability,liu2023trustworthy}, and saliency maps (for visual data~\citep{Adebayo2018SanityCF}); 
\textit{ii}) 
X-of-Thought approaches that break down complex tasks into structured reasoning steps
to make the reasoning process more interpretable to users, \textit{e}.\textit{g}., chain-of-thought~\citep{wei2022chain,zhang2022automatic,diao2023active,fu2023chain}, tree-of-thought~\citep{long2023large,yao2024tree,cao2023probabilistic}, graph-of-thought~\citep{besta2024graph,lei2023boosting}, and their variants; 
\textit{iii})
embedding watermarks to enhance traceability and content provenance for AI-generated outputs~\citep{yu2023leaked,yu2023safe,SynthID_dathathri2024scalable,SynthID}; 
\textit{iv}) leveraging external knowledge sources, tools, or methods as reference for reasoning, \textit{e}.\textit{g}., retrieval-augmented generation (RAG)~\citep{jiang2023active,lewis2020retrieval}, function calling~\citep{erdogan2024tinyagent,lin2024hammer,ran2024alopex}, and web browsing~\citep{nakano2021webgpt}; 
\textit{v}) leveraging interpretability algorithms or models as auxiliary tools to generate insights into AI outputs while automating AI interpretability~\citep{ribeiro2016should,ribeiro2016model,bills2023language}.

%% file: src/ai_safety_practices_table.tex
\begin{table*}[t]
\centering
\begin{tabular}{@{}p{2.5cm}|p{4.6cm}|p{8.6cm}@{}}
\toprule
\textbf{Topic} & \textbf{Research}  & \textbf{Industry Practices} \\ \midrule
\multirow{3}{10em}{\textbf{Jailbreaking}} & \cite{hubinger2024sleeper,chao2023jailbreaking,zou2023universal,wu2024new,Liu2023Jailbreaking,liu2023prompt,pedro2023prompt,choquet2024exploiting,piet2024jatmo,perez2022ignore,suo2024signed,toyer2023tensor,liu2023query,wei2024jailbroken,diaa2024optimizing,Shen2023DoAN,song2024multilingual,yan2024llm,teng2024heuristic}, etc.& \multirow{3}{10em}{PromptArmor~\cite{PromptArmor}} \\\midrule
\multirow{2}[2]{*}{\textbf{Red teaming}}
 & \multirow{3}{14em}{\cite{ganguli2022red,wallace2019trick,ribeiro-etal-2020-beyond,bartolo2021improving,perez2022red,longpre2024safe,yu2024llm,tedeschi2024alert,mazeika2024harmbench,derczynski2024garak}, etc.}  & Purple Llama~\cite{bhatt2023purple}, DeepMind~\cite{perez2022red}, Anthropic~\cite{ganguli2022red}, Protect AI~\cite{protect_ai}, Giskard~\cite{giskard}, Virtue AI~\cite{VirtueAI}, Dynamo AI~\cite{dynamoai}, Mindgard~\cite{mindgard}, etc.\\\midrule
\multirow{4}{10em}{\textbf{Moderation}} & \multirow{4}{14em}{\cite{kumar2024watch,wu2024legilimens,qiao2024scaling,elesedy2024lora,ma2023adapting,jha2024memeguard,peng2019transfer,xu2020recipes,chen2024finding,kumar2023certifying}, etc. } & OpenAI Moderation API~\cite{openai-data-paper}, Perspective API~\cite{perspective-api}, Detoxify~\cite{Detoxify}, Llama Guard~\cite{inan2023llamaguard}, Emergence AI~\cite{niknazar2024building}, Giskard~\cite{giskard}, Dynamo AI~\cite{dynamoai}, Calypso AI~\cite{calypsoai}, Lakera AI~\cite{lakera}, BreezeML~\cite{breezeML}, etc.\\\midrule
\textbf{Customized Workflow} & \multirow{2}{10em}{-} & Guardrails AI~\cite{guardrails}, Nvidia Nemo Guardrails~\cite{nemo}, Protect AI~\cite{protect_ai}, etc.\\\midrule
\multirow{3}{10em}{\textbf{Privacy}} & \multirow{3}{14em}{\cite{duan2023privacy,li2024llm,miranda2024preserving,carlini2021extracting,zhu2019deep,hu2022membership,zhang2024understanding,sebastian2023privacy}, etc.} & Dynamo AI~\cite{dynamoai}, ProtectAI~\cite{protect_ai}, Private AI~\cite{private_ai}, PromptArmor~\cite{PromptArmor}, DataGrail~\cite{datagrail}, Dastra~\cite{dastra}, OneTrust~\cite{onetrust}, Relyance AI~\cite{relyanceAI}, Zendata~\cite{zendata}, etc.\\\midrule
\multirow{3}{10em}{\textbf{Defenses}} & \cite{Robey2023SmoothLLM,huang2024safealigner,Varshney2023TheAO,phute2023llm,wu2024llms,xiong2024defensive,openai-data-paper,cao2024nothing,fu2024cross,wang2024selfdefend,Zeng2024AutoDefenseML,huang2024safealigner,papernot2016distillation,gowal2018effectiveness,cohen2019certified,du2024mogu,shi2022just,shi2021selective}, etc.& OpenAI Rule-based rewards~\cite{murule}; DeepMind Weight Averaged
Reward Models~\cite{rame2401warm}; others see \textit{Moderation}, \textit{Guardrail}, and \textit{Privacy}\\\midrule
\textbf{Serving Security} &\cite{wu2024new} 
 &ProtectAI~\cite{protect_ai}, Relyance AI~\cite{relyanceAI}, Transcend~\cite{Transcend}, etc.\\\midrule
\multirow{5}{10em}{\textbf{AI governance}} & \multirow{5}{14em}{\cite{yu2023leaked,yu2023safe,alabdulakreem2024securellm,kirchenbauer2023watermark,ehsan2021expanding,chao2024watermarking}, etc.} & Giskard~\cite{giskard}, ProtectAI~\cite{protect_ai}, Calypso AI~\cite{calypsoai}, Saidot~\cite{saidot}, Arize~\cite{arize}, Dynamo AI~\cite{dynamoai}, Credo AI~\cite{credo_ai}, Google~\cite{raji2020closing}, DataGrail~\cite{datagrail}, Dastra~\cite{dastra}, OneTrust~\cite{onetrust}, Relyance AI~\cite{relyanceAI}, Transcend~\cite{Transcend}, Zendata~\cite{zendata}, DeepMind SynthID watermark~\cite{SynthID_dathathri2024scalable,SynthID}, etc.\\ \bottomrule
\end{tabular}
\caption{Overview of Research and Industry Practice in AI Safety Topics}
\label{tab:ai-safety-practice}
\end{table*}

%% file: chapters/7_ai_safety_practitioners_in_2020s.tex
\section{AI Safety in the 2020s: Challenges and Missions}\label{sec:ai_safety_missions}

This section utilizes the blueprint to guide our efforts at the current stage. As AI safety practitioners in the 2020s, we stand at a pivotal moment in history, with society moving toward an era where AI will be deeply integrated into daily life. While addressing immediate safety concerns, our primary mission is to anticipate and mitigate risks within the expanding AI landscape. Thus, certain directions may require attention.

\noindent\textbf{A shift from AI safety to AI quality assurance.} The rapid changing AI landscape requires a shift of our focus from narrowly defined AI safety concerns to a more comprehensive concept, \textit{i}.\textit{e}., \textit{AI quality assurance}.
Besides focusing on safety and security, practitioners can put more efforts on addressing alignment~\citep{ouyang2022training} as well as the quality of interactions between users and AI. 
Viewing \textit{AI safety} as part of \textit{AI quality assurance} helps enhance AI in a way that is safe, accurate, and capable of meeting the complex requirements of real-world AI applications. The terminology ``AI quality assurance'' also guides us to think more about what we can do to align our current efforts with the blueprint picture.

\noindent\textbf{Enhancing alignment with more interdisciplinary insights. }
Aligning AI with deeper insights across different fields, people, and cultures is important for enhancing the services provided by AI models. Current AI models, while advanced, often lack a deep understanding of ethical, social, and cultural contexts, while most of the existing work on alignment tend to be conducted by people in the computer science field. 
However, more advanced AI systems necessitate interdisciplinary collaborations between AI researchers and experts from diverse fields such as psychology, sociology, history, art, and anthropology, especially for AI applications that involve frequent interactions with different user groups, such as domains like finance~\citep{wu2023bloomberggpt,zhang2023privacy}, healthcare~\citep{thirunavukarasu2023large,abbasian2023conversational}, education~\citep{rahman2023chatgpt,mbakwe2023chatgpt}, etc. 
AI safety practitioners might expand their focus beyond technical aspects and foster interdisciplinary collaborations, such that AI systems can be technically robust while being socially and ethically aligned with diverse user backgrounds.

\noindent\textbf{Hallucination.} 
Hallucination remains a fundamental and persistent challenge and is difficult to eliminate entirely, regardless of advancements in AI~\citep{huang2023survey}. 
This is because hallucinations can originate at nearly every stage of AI lifecycle. Factors such as the quality and representativeness of training data, inappropriate user inputs, and inference processes all contribute to the occurrence hallucinations. 
Despite implementing RAG~\citep{lewis2020retrieval} helps mitigate hallucination, it remains an inherent risk of AI models. Thus, implementing a comprehensive moderation pipeline for identifying and addressing hallucinations is crucial, which involves grounding user queries with RAG, detecting presence of hallucinations, and fixing hallucinations in the model outputs if possible.

\noindent\textbf{Customization. }
Customizing AI workflows is essential to ensure that AI systems address diverse user needs effectively and align with specific cultural norms, legal regulations, and ethical considerations.
Current AI systems such as ChatGPT~\citep{openai-data-paper} tend to provide similar or identical services to all user groups, regardless of individual differences. 
However, as AI becomes increasingly integrated into daily life, one-size-fits-all solutions are inadequate. 
With diverse backgrounds and needs, users might expect AI systems to adapt to their individual differences and preferences, and deliver safe, effective, and personalized services.
As an example, AI models interacting with children should respond differently compared with interacting with adults, necessitating considering user age, learning objectives, cognitive abilities, emotional maturity, etc. 
In domains 
such like finance~\citep{wu2023bloomberggpt,zhang2023privacy}, healthcare~\citep{thirunavukarasu2023large,abbasian2023conversational}, and  education~\citep{rahman2023chatgpt,mbakwe2023chatgpt}, 
customizing AI services based on user profiles is crucial for enhancing safety 
while improving user experiences.

\noindent\textbf{Interaction. }
Our long-term vision for the blueprint future suggests that the \textit{Internet of Everything} involves complex interactions between intelligent robots and human users (similar ideas also discussed in~\citep{bajcsy2024human,gabriel2024ethics}), which calls for a shift in our focus from static, one-shot analyses to dynamic, context-aware interactions between users and models to improve alignment.
Beyond prompt-based AI systems and services, we can develop more sophisticated, context-aware interfaces for human-AI interactions to support multi-round interactions that adapt to evolving conversational contexts.
Also, real-time safety protocols are in need to adjust to changing contexts and user needs. 
It is also essential to develop methodologies for safety assessment that functions effectively in complex and rapidly evolving settings, such as those involving interactions between multiple AI agents and human users.

%% file: chapters/8_conclusion.tex
\section{Conclusion}\label{sec:conclusion}
This paper presents a blueprint for an advanced human society and leverages this forward-looking vision to guide today’s AI safety efforts. 
Through the blueprint, it becomes clear that Artificial General Intelligence (AGI) is not the ultimate goal of AI development. Instead, the true vision 
lies in the \textit{Internet of Everything}, a deeply interconnected world where intelligent systems seamlessly integrate into daily life.
While AGI is a popular topic discussed more among people in the computer science field today, a more advanced world demands interdisciplinary collaborations across various fields. Regarding AGI as the ultimate goal might limit our creativity in the 2020s, trapping us in ``local optima'' and potentially causing us to overlook the real challenges.

What will this future world be like? Will it bring more happiness to human-beings? The answer is uncertain. Advanced technologies may take on more duties of daily life, leaving humans with fewer tasks and more predictability.
On the other hand, with more daily problems being solved by technologies, people may have more time for self-reflection and personal growth, which may lead to another form of ``happiness''.

As we stand at a pivotal moment in history, our efforts become more than an incremental technological endeavor, but a profound exploration of human potential and the boundaries of human capability.
In any case, the journey toward this advanced future is fascinating, and every human being will look forward to it.